\documentclass[aps,prev,twocolumn,preprintnumbers,floatfix,nofootinbib]{revtex4-1}
\pdfoutput=1

\usepackage{graphicx}
\usepackage{epstopdf}
\usepackage{mathrsfs}
\usepackage{amssymb}
\usepackage{verbatim}
\usepackage{color}
\usepackage{multirow}
\usepackage{amsmath}
\usepackage{slashed} 
\usepackage{float}
\usepackage[normalem]{ulem}
\usepackage{sidecap}
\usepackage{hyperref}
\usepackage{cancel}
\usepackage{enumerate}
\hypersetup{pdfstartview=FitV,colorlinks=true,linkcolor=blue,citecolor=red,filecolor=black,urlcolor=blue}

\def\stacksymbols #1#2#3#4{\def\theguybelow{#2}
    \def\vp{\lower#3pt}
    \def\sp{\baselineskip0pt\lineskip#4pt}
    \mathrel{\mathpalette\intermediary#1}}

\def\intermediary#1#2{\vp\vbox{\sp
     \everycr={}\tabskip0pt
     \halign{$\mathsurround0pt#1\hfil##\hfil$\crcr#2\crcr
              \theguybelow\crcr}}}

\newcommand{\gev}{\, {\rm GeV}}

\newcommand{\beq}{\begin{equation}}
\newcommand{\eeq}{\end{equation}}
\newcommand{\bea}{\begin{eqnarray}}
\newcommand{\eea}{\end{eqnarray}}

\newcommand{\gsim}{\lower.7ex\hbox{$\;\stackrel{\textstyle>}{\sim}\;$}}
\newcommand{\lsim}{\lower.7ex\hbox{$\;\stackrel{\textstyle<}{\sim}\;$}}

\newcommand{\trh}{T_{\mathrm{RH}}}

\newcommand{\lslashslash}{%

  \raisebox{0.8ex}{%
    \scalebox{.7}{%
      \rotatebox[origin=c]{18}{$-$}%
    }%
  }%
}

\newcommand{\lslash}{%
  {%
   \vphantom{d}%
   \ooalign{\kern-.1em\smash{\lslashslash}\hidewidth\cr${\rm l}$\cr}%
   \kern.05em
  }%
}

\def\be{\begin{equation}}
\def\ee{\end{equation}}
\def\bea{\begin{eqnarray}}
\def\eea{\end{eqnarray}}

\def\sp{\;\;\;,\;\;\;}

\def\lsim{\raise0.3ex\hbox{$\;<$\kern-0.75em\raise-1.1ex\hbox{$\sim\;$}}}
\def\gsim{\raise0.3ex\hbox{$\;>$\kern-0.75em\raise-1.1ex\hbox{$\sim\;$}}}

\def\inbar{\,\vrule height1.5ex width.4pt depth0pt}

\def\IC{\relax\hbox{$\inbar\kern-.3em{\rm C}$}}
\def\IQ{\relax\hbox{$\inbar\kern-.3em{\rm Q}$}}
\def\IR{\relax{\rm I\kern-.18em R}}
 \font\cmss=cmss10 \font\cmsss=cmss10 at 7pt
\def\IZ{\relax\ifmmode\mathchoice
 {\hbox{\cmss Z\kern-.4em Z}}{\hbox{\cmss Z\kern-.4em Z}}
 {\lower.9pt\hbox{\cmsss Z\kern-.4em Z}}
 {\lower1.2pt\hbox{\cmsss Z\kern-.4em Z}}\else{\cmss Z\kern-.4em Z}\fi}

\def\rhorh{\rho_{\rm RH}}

\def\ae{a_{\rm end}}
\def\rhoe{\rho_{\rm end}}

\def\be{\begin{equation}}
\def\ee{\end{equation}}
\def\bea{\begin{eqnarray}}
\def\eea{\end{eqnarray}}

\def\trh{T_{\rm RH}}

\def\comment#1{}

\def\u1x{U(1)_X}
\newcommand{\nc}{\newcommand}
\nc{\LL}{L}
\nc{\vv}{\tilde{v}}
\nc{\ccdot}{\!\cdot\!}
\nc{\gsm}{G_{SM}}
\nc{\vfive}{\mathbf{5}\oplus\mathbf{\overline{5}}}
\nc{\vten}{\mathbf{10}\oplus\mathbf{\overline{10}}}
\nc{\zhol}{Z^{\rm hol}}
\nc{\xfb}{\,{\rm fb}}

\begin{document}

\preprint{CERN-TH-2022-071}
\preprint{UMN--TH--4121/22}
\preprint{FTPI--MINN--22/12}

\vspace*{1mm}

\title{Inflationary Gravitational Leptogenesis}

\author{Raymond T. Co$^{a}$}
\email{rco@umn.edu}
\author{Yann Mambrini$^{b}$}
\email{yann.mambrini@th.u-psud.fr}
\author{Keith A. Olive$^{a}$}
\email{olive@physics.umn.edu}

\vspace{0.1cm}

 \affiliation{
${}^a$ 
 William I.~Fine Theoretical Physics Institute, 
       School of Physics and Astronomy,
            University of Minnesota, Minneapolis, MN 55455, USA
}
\affiliation{
${}^b$ Universit\'e Paris-Saclay, CNRS/IN2P3, IJCLab, 91405 Orsay, France
 }

\begin{abstract} 

We consider the generation of the baryon asymmetry in models with right-handed neutrinos produced through gravitational scattering of the inflaton during reheating. The right-handed neutrinos later decay and generate a lepton asymmetry, which is partially converted to a baryon asymmetry by Standard Model sphaleron processes. We find that a sufficient asymmetry can be generated for a wide range of right-handed neutrino masses and reheating temperatures. We also show that the same type of gravitational scattering produces Standard Model Higgs bosons, which can achieve inflationary reheating consistent with the production of a baryon asymmetry.

\end{abstract}

\maketitle

{\bf \textit{Introduction}}.---%
One of the most elegant mechanisms for generating the baryon asymmetry of the Universe is leptogenesis \cite{FY}. In its simplest version, the lepton number violating out-of-equilibrium decay of a heavy right-handed neutrino produces a lepton asymmetry if C and CP are violated in the decays. Then the baryon number $B$ and lepton number $L$ violating (but $B-L$ conserving) sphaleron processes \cite{sphal} distribute the asymmetry between leptons and baryons. 
As long as the effective lepton number violating operators
remain out of equilibrium, the baryon (and lepton) asymmetry will be preserved \cite{FY2,HT,Nelson:1990ir}. 

The obvious attractiveness in models of leptogenesis,
is the fact that the only element beyond the Standard Model required is a massive right-handed neutrino,
often introduced to generate light Standard Model neutrino masses via the seesaw mechanism \cite{seesaw}. Differences among models of leptogenesis often relate to the means by which right-handed neutrinos populate the Universe or their embedding in a UV completion of the Standard Model \cite{leptogenesis}. 
For example, a common assumption is that the right handed neutrinos are in thermal equilibrium in the radiation bath newly created after inflation \cite{thermal}. This requires that the reheating temperature $\trh$ exceeds the right-handed neutrino mass $m_N$. However, this requirement is not necessary, as right-handed neutrinos produced in the decay of the inflaton, provide a direct source of non-thermal right-handed neutrinos \cite{Giudice:1999fb} and require only $m_N < m_\phi/2$, where $m_\phi$ is the mass of the inflaton $\phi$.

Most models of non-thermal leptogenesis carry some form of additional model dependence, namely, how one couples the inflaton to the right-handed neutrino. In some cases, one might associate the inflaton with the supersymmetric partner of the right-handed neutrino \cite{eno8}, or in supergravity models there maybe a gravitational coupling induced by the chosen
forms of K\"ahler potential and superpotential \cite{egno4,kmov}. 

In this \emph{Letter}, we consider a model-independent formulation for the production of right-handed neutrinos leading to leptogenesis. 
That is, once the the inflaton potential is specified, we make no assumptions about how the inflaton couples to the right-handed neutrino sector. Its production from the inflaton condensate is purely gravitational\footnote{Note that production from the inflaton condensate almost always dominates over the gravitational production of matter from the thermal bath \cite{garny}.}  \cite{ema,MO,Barman:2021ugy,CMOV}. We consider as an example, the class of inflationary models called T-models \cite{Kallosh:2013hoa} and show that for all such models for which the equation of state parameter during the period of reheating $w \ge 0.5$, the proper baryon asymmetry may be generated for reasonable choice of the right-handed neutrino mass and reheating temperature.

In what follows, we first compute the number density of right-handed neutrinos produced gravitationally from inflaton oscillations. 
We then apply this result to obtain the resulting
baryon/lepton asymmetries from the out-of-equilibrium decay of the right-handed neutrinos. 
We also show that even if 
the Higgs-inflaton coupling is only gravitational
(minimal or non-minimal),
it may be possible to {\it simultaneously} produce the lepton asymmetry as well as the entropy of the universe without the need to consider a specific coupling of the inflaton to matter.

{\bf \textit{Gravitational Production Rates}}.---%
The simplest process for producing a lepton asymmetry from the out-of-equilibrium decay of a right-handed neutrino is 
a direct decay of the inflaton to $N$. If such a coupling exists and $m_{N}$ exceeds the maximum temperature after inflation, then $N$ will be produced out-of-equilibrium thus realizing the original leptogenesis scenario \cite{FY}. 
However, even in the absence of a direct coupling between the inflaton and the right-handed neutrino, $N$ can be produced from scattering within the thermal bath or 
directly from the inflaton condensate. The former is the common mechanism leading to thermal leptogenesis \cite{thermal}. The latter, on the other hand, is inevitable for 
processes mediated by gravity \cite{MO,CMOV}.

\begin{figure}[ht]
\centering
\includegraphics[width=2.5in]{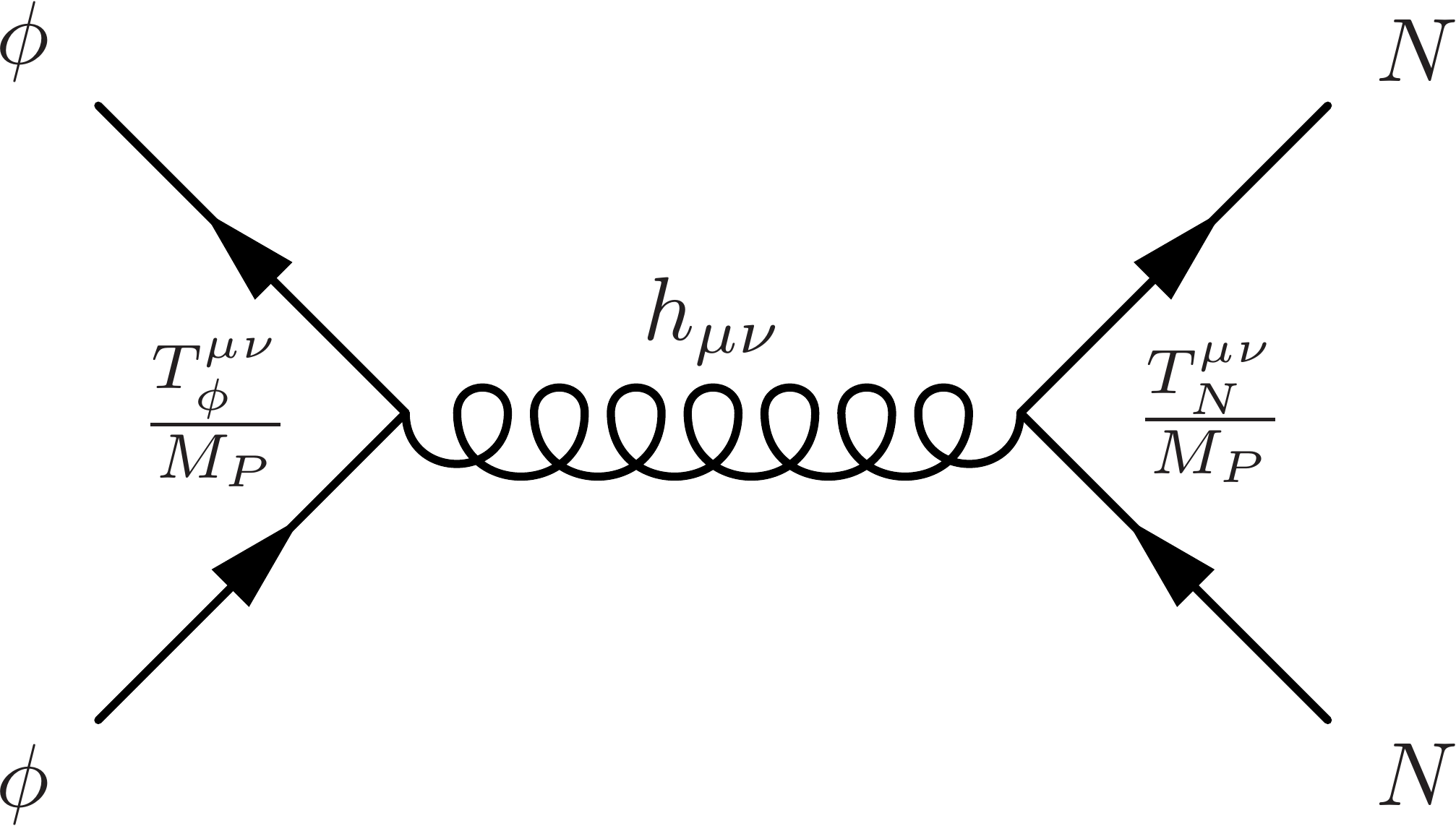}
\caption{\small Feynman diagram for the production of right-handed neutrinos $N$ through the gravitational scattering of the inflaton condensate $\phi$.
}
\label{Fig:feynman}
\end{figure}

Gravitational interactions are described by the Lagrangian~(see e.g., \cite{hol})
\beq
\sqrt{-g}{\cal L}_{\rm int}= -\frac{1}{M_P}h_{\mu \nu}
\left(T^{\mu \nu}_{\rm SM}+T^{\mu \nu}_\phi + T^{\mu \nu}_{N} \right) \, .
\label{Eq:lagrangian}
\eeq
Here SM represents Standard Model fields, $\phi$ is the inflaton and $N$ is the right-handed neutrino. We assume the standard form of the stress-energy tensor $T^{\mu \nu}_i$ which depends on the spin of the field, $i = 0$, $1/2$, $1$.  In Fig.~\ref{Fig:feynman}, we show the $s$-channel exchange of a graviton obtained from the Lagrangian~(\ref{Eq:lagrangian}) for the production of right-handed neutrinos from  the inflaton condensate. In addition, a similar diagram exists
for the production of Standard Model fields during the reheating process. 
The Planck suppression due to graviton exchange is partially compensated by the energy available in the inflaton condensate at the end of inflation.

The rate for producing right-handed neutrinos can be found in Refs.~\cite{CMOV,Bernal:2021kaj} where it has been calculated in the context of a fermionic dark matter candidate. Because the production mechanism depends on the time-dependent oscillations of the inflaton subsequent to the period of exponential expansion, 
we must specify examples of the inflaton potential we consider.

As a specific example, we consider the $\alpha$-attractor T-model~\cite{Kallosh:2013hoa}, 
\begin{equation}
    V(\phi) \; = \;\lambda M_P^{4}\left|\sqrt{6} \tanh \left(\frac{\phi}{\sqrt{6} M_P}\right)\right|^{k} \, ,
\label{Eq:Vatt}
\end{equation}
which can in fact arise from a simple superpotential \cite{GKMO1}
\begin{equation}
    W=2^{\frac{k}{4}+1} \sqrt{\lambda} M_{P}^3 \left(\frac{(\phi/M_{P})^{\frac{k}{2}+1}}{k+2}-\frac{(\phi/M_{P})^{\frac{k}{2}+3}}{3(k+6)}\right) ,
\end{equation}
in the context of no-scale supergravity. 
Eq.~(\ref{Eq:Vatt}) can be expanded about the origin
\begin{equation}
    \label{Eq:potmin}
    V(\phi)= \lambda \frac{\phi^{k}}{M_{P}^{k-4}}, \quad {\rm for}\ \phi \ll M_{P} \, ,
\end{equation}
and we will use this generic form hereafter. Therefore, it should be noted that the remaining discussion is general and not limited to T-models of inflation. Phenomenological aspects of T-models were recently considered in~\cite{egnov}.

At the end of inflation, the time-dependent oscillating inflaton field can be parametrized as
\begin{equation}
    \label{Eq:oscillation}
    \phi(t) \; = \; \phi_0(t) \cdot \mathcal{P}(t) \, , 
\end{equation}
where $\phi_0(t)$ is the 
amplitude and 
\beq
\mathcal{P}(t)=\sum_{n=-\infty}^{n=+\infty}{\cal P}_n e^{-in\omega t}
\eeq
describes the periodicity of the oscillations, with 
frequency $\omega$
given by \cite{GKMO2}
\beq
\label{eq:angfrequency}
\omega=m_\phi \sqrt{\frac{\pi k}{2(k-1)}}
\frac{\Gamma(\frac{1}{2}+\frac{1}{k})}{\Gamma(\frac{1}{k})}\, ,
\eeq
where $m^2_\phi=V''(\phi_0)$.
The rate (per unit time and volume) for the production of right-handed neutrinos can be expressed as~\cite{CMOV}
\beq
\label{eq:rateferm}
R^{\phi^k}_{N}=\frac{2 \times \rho_\phi^2}{4 \pi M_P^4}
\frac{m_N^2}{m_\phi^2}
\Sigma_{1/2}^k \, ,
\eeq
where during oscillations $\rho_\phi=V(\phi_0)$ is the energy density of the inflaton
and the explicit factor of two accounts for two particles produced in the final state.  
Considering the potential (\ref{Eq:potmin}), 
the effective inflaton mass squared is given by $m_\phi^2 = \lambda k(k-1) \phi_0^{k-2}/M_P^{k-4}$ and
\begin{equation}
    \label{eq:ratefermion2}
    \Sigma_{1/2}^k = \sum_{n=1}^{+\infty} |{\cal P}_n^k|^2
    \frac{m_\phi^2}{E_n^2}
\left[1-\frac{4 m_N^2}{E_n^2}\right]^{3/2} \, ,
\end{equation}
where the Fourier modes ${\cal P}_n^k$ are obtained from
 \cite{Ichikawa:2008ne,Kainulainen:2016vzv,GKMO2}
\beq
V(\phi)=V(\phi_0)\sum_{n=-\infty}^{\infty} {\cal P}_n^ke^{-in \omega t}
=\rho_\phi\sum_{n = -\infty}^{\infty} {\cal P}_n^ke^{-in \omega t} \, ,
\eeq
$E_n = n \omega$ being the energy of the $n$-th inflaton oscillation mode. The values of $\Sigma_{1/2}^k$ are given in Table~\ref{tab:Sigma_k} for $k \le 20$.

\begin{table}[ht!]
\begin{tabular}{|l|l|l|l|}
\hline
$\Sigma_{1/2}^6 = 0.101 $
    & $\Sigma_{1/2}^8 = 0.133 $
    & $\Sigma_{1/2}^{10} = 0.157 $
    & $\Sigma_{1/2}^{12} = 0.177 $ \\ \hline
$\Sigma_{1/2}^{14} = 0.192 $
    & $\Sigma_{1/2}^{16} = 0.205 $
    & $\Sigma_{1/2}^{18} = 0.216 $
    & $\Sigma_{1/2}^{20} = 0.225 $ \\ \hline
\end{tabular}
\caption{\small Coefficients $\Sigma_{1/2}^k$ relevant for the gravitational production rate of right-handed neutrinos.}
\label{tab:Sigma_k}
\end{table}

The number density of right-handed neutrinos, $n_N$,
is obtained by solving the Boltzmann equation, which can be expressed as
\beq
\frac{d (n_N a^3)}{da}=\frac{a^2R^{\phi^k}_{N}(a)}{H} \, ,
\label{Eq:boltzmann2}
\eeq
where $a$ is the cosmological scale factor. The $a$ dependence in the rate comes from the 
evolution of $\rho_\phi$, which is given by  \cite{GKMO1,GKMO2}
\beq
\rho_\phi(a) = \rho_{\rm end} \left(\frac{a_{\rm end}}{a} \right)^\frac{6k}{k+2} \, ,
\label{rhophia}
\eeq
where $a_{\rm end}$ is the scale factor when inflation ends (defined when the slow-roll parameter equals unity). The equation of state for $\phi$ is $w = (k-2)/(k+2)$.

Eq.~(\ref{Eq:boltzmann2}) for the density of right-handed neutrinos can be solved analytically
 \cite{CMOV},
\bea
&&
n_{N}(a_{\rm RH})
\simeq 
\frac{m_N^2\sqrt{3}(k+2)\rhorh^{\frac{1}{2}+\frac{2}{k}}}{12 \pi k(k-1)\lambda^{\frac{2}{k}}M_P^{1+\frac{8}{k}}}
\left(\frac{\rhoe}{\rhorh}\right)^{\frac{1}{k}}
\Sigma_{1/2}^k \ \ \ \ 
\label{Eq:nhalf}
\eea
evaluated at the time of reheating, which for now we assume is a result of the decay of the inflaton to Standard Model particles. We define
$\rhorh$ as the energy density in radiation when it becomes equal to the inflaton energy density and $\rhorh = (\pi^2 g_*(\trh)/30) \trh^4$, with $g_*(\trh)$ the number of relativistic degrees of freedom\footnote{$g_* = 427/4$ for the full Standard Model particle content.} at $\trh$.

{\bf \textit{Leptogenesis}}.---%
Once produced, the right-handed neutrinos decay rapidly,
\begin{equation}
\begin{split}
& N \rightarrow L_{\alpha} + H \\
& N \rightarrow \bar{L}_{\alpha} + \overline{H} ,
\end{split}
\end{equation}
where $L$ and $H$ are the left-handed lepton and Higgs electroweak doublets respectively.
If CP is violated in the decay of $N$, then a lepton asymmetry 
\begin{equation}
Y_L \equiv \frac{n_L}{s} = \epsilon \frac{n_{N}}{s} 
\end{equation}
is produced. Here $s = (2 \pi^2 g_*(\trh) /45) \trh^3$
is the entropy density. The CP violation is encapsulated in \cite{luty,CPviol}
\begin{equation}
\label{CP1}
\epsilon \equiv \frac{\Gamma_{N \rightarrow L_{\alpha} H} - \Gamma_{N \rightarrow \bar{L}_{\alpha} \overline{H}}}{\Gamma_{N \rightarrow L_{\alpha} H} + \Gamma_{N \rightarrow \bar{L}_{\alpha} \overline{H}}} .
\end{equation}
A non-zero value for $\epsilon$ requires
at least two right-handed neutrinos. We assume the existence of three right-handed neutrinos and
denote the lightest of these as $N$ with mass $m_N$. The remaining two will be denoted as $N_{2,3}$ with masses $m_{2,3}$ and we assume $m_N \lesssim m_\phi \ll m_{2, \, 3}$. Furthermore, we assume that the light and mostly left-handed neutrino masses are determined by the seesaw mechanism \cite{seesaw} so that 
\begin{equation}
\label{neuti}
m_{\nu_i} \simeq \frac{|y_{i}|^2 v^2}{m_i},
\end{equation}
where $y_i$ is a Yukawa coupling, 
and $v \approx 174$ GeV is the Standard Model Higgs expectation value. Using the seesaw expression, 
we can write \cite{kmov}
\begin{equation}
\label{cpneut}
\epsilon \simeq -\frac{3 \, \delta_{\text{eff}}}{16 \pi} \cdot \frac{m_{\nu_i} \, m_N}{v^2}, 
\end{equation}
where $\delta_{\text{eff}}$ is the effective CP violating phase in the neutrino mass matrix and $0\leq \delta_{\rm eff}\leq 1$.

Finally, this lepton asymmetry is converted into the baryon asymmetry via the electroweak sphaleron processes that freeze out at the electroweak phase transition, giving $Y_B = \frac{28}{79} Y_L$~\cite{spha2,HT,kmov} and 
\begin{equation}
Y_B \simeq 3.5 \times 10^{-4} \delta_{\text{eff}} \frac{n_{N}}{s} \left(\frac{m_{\nu_i}}{0.05~{\rm eV}} \right) \left( \frac{m_N}{10^{13}~{\rm GeV}} \right) \, ,
\end{equation}
while the observed value is $Y_B \simeq 8.7 \times 10^{-11}$~\cite{Planck:2018jri}.

\begin{figure}[t]
\centering
\includegraphics[width=\linewidth]{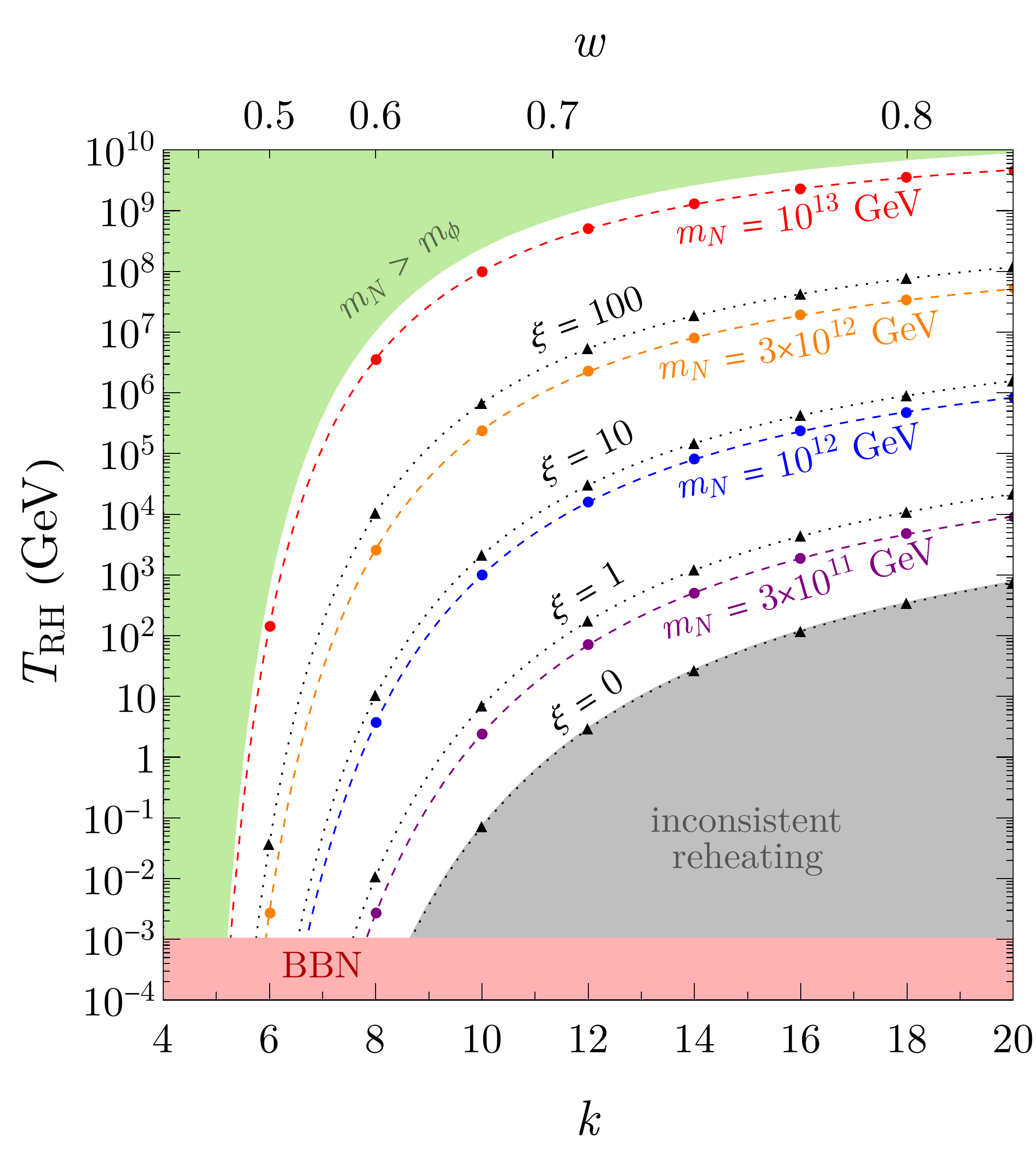}
\caption{\small The colored dashed curves show values of the reheating temperature $\trh$ required to explain the observed baryon asymmetry as a function of $k$, the exponent of the potential defined in Eq.~(\ref{Eq:potmin}), for different choices of the right-handed neutrino mass $m_N$. The black dotted lines show the reheating temperature obtained from purely gravitational reheating for different choices of $\xi$, the non-minimal gravitational coupling constant defined in Eq.~(\ref{Eq:couplingricci}). A minimal gravitational coupling ($\xi = 0$) gives a minimum $T_{\rm RH}$, excluding the gray region. The upper x-axis labels the equation of state $w = (k-2)/(k+2)$ of the inflaton, during reheating.}
\label{Fig:baryon_asymmetry_k}
\end{figure}

\begin{figure}[t]
\centering
\includegraphics[width=\linewidth]{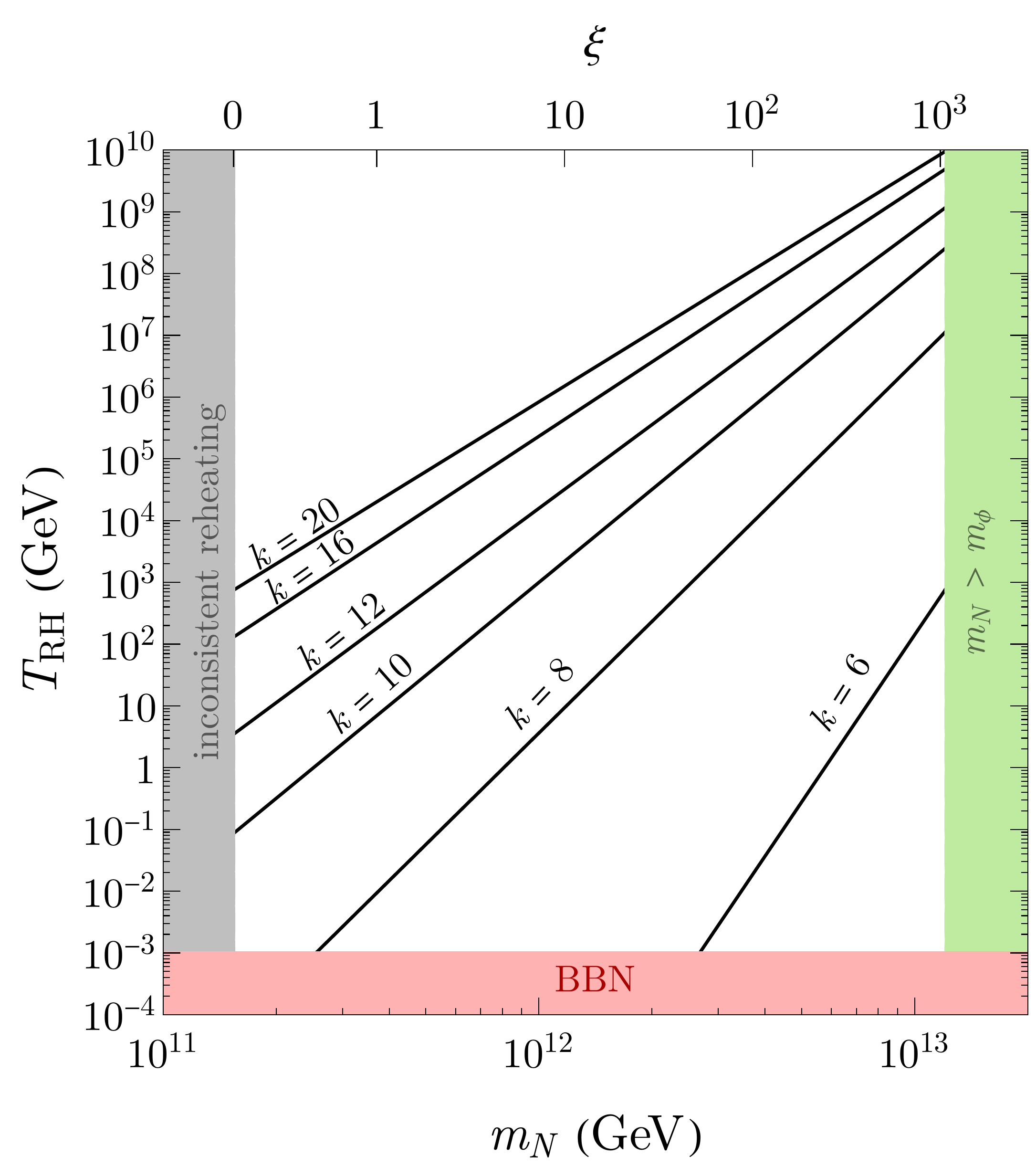}
\caption{\small Values of the reheating temperature $\trh$ required to explain the observed baryon asymmetry as a function of $m_N$ for different choices of $k$. The upper x-axis labels the approximate values of $\xi$ necessary to achieve the corresponding $T_{\rm RH}$ through purely gravitational reheating.}
\label{Fig:baryon_asymmetry_mN}
\end{figure}

The required reheating temperature $\trh$ for specific choices of the right-handed neutrino mass $m_N$ as a function of the equation of state parameter $w$ is displayed in Fig.~\ref{Fig:baryon_asymmetry_k}. Here we take $\rho_{\rm end} = (5 \times 10^{15} \gev)^4$ and $\lambda = 18\pi^2 A_{S*}/(6^{k/2} N_*^2)$ with $A_{S*}$ being the amplitude of the curvature power spectrum measured to be $\ln(10^{10} A_{S*}) = 3.044$~\cite{Planck:2018jri,Planck:2018vyg} and the number of $e$-folds $N_* = 55$ for the Planck pivot scale $k_* = 0.05$ Mpc$^{-1}$.%
\footnote{The variables $\rho_{\rm end}$, $\lambda$, and $N_*$ are in principle functions of $k$ and $\trh$~\cite{GKMO2} in order to explain the CMB observations, but fixing to the aforementioned values gives an excellent approximation.}
This choice of parameters leads to an inflaton mass, $m_\phi \simeq 1.2 \times 10^{13} \gev$, at the end of inflation which places an upper bound on $m_N$ due to kinematics. The circles along each curve correspond to even values of $k$ in the inflaton potential. For $k = 2$,  from Eq.~(\ref{Eq:nhalf}), $n_N/s \propto m_N^2 \trh$ and $Y_B \simeq 10^{-69} m_N^3 \trh/{\rm GeV}^4$ is far too small to produce the required asymmetry. For $k=4$, $Y_B$ is independent of $\trh$ but requires a large $m_N \simeq 3\times10^{14}$ GeV, which exceeds the inflaton mass.  However, for $k\ge 6$, for reasonable $\trh$ and a sufficiently large $m_N$, the proper asymmetry can be generated. This is also demonstrated in Fig.~\ref{Fig:baryon_asymmetry_mN}, where the required $\trh$ as a function of $m_N$ is shown for different values of $k$. For $k=6$, the correct asymmetry is obtained for $T_R \simeq 10^2 \gev (m_N/10^{13}\gev)^9$. More generically, $T_R \propto m_N^{3k/(k-4)}$ for any $k \neq 4$. Lastly, the green regions are not feasible because the gravitational production of $N$ is kinematically forbidden. 

{\bf \textit{Gravitational reheating}}.---%
The previous calculation of the lepton asymmetry, was based solely on gravity for the production of right-handed neutrinos and $n_L$. However, we
assumed that the entropy (and hence $\trh$) was produced by the decay of the inflaton to radiation. For example, a coupling $y \phi {\bar f} f$ would produce a reheating temperature proportional to $y^{k/2}$ for $k<7$ and to $y^{3k/(2k-8)}$ for $k>7$ \cite{GKMO2} and would allow temperatures in the range shown in Fig.~\ref{Fig:baryon_asymmetry_k} for $y<1$.
We now ask whether both the numerator and denominator in $Y_L = n_L / s$ can be produced purely by gravity. 
It was recently shown that a gravitational coupling could be sufficient to ensure a complete reheating of the Universe \cite{CMOV}. However, to ensure a sufficiently large reheating temperature ($\trh \gtrsim 1$ MeV) to avoid conflict with Big Bang nucleosynthesis, which excludes the red regions in Figs.~\ref{Fig:baryon_asymmetry_k} and \ref{Fig:baryon_asymmetry_mN}, one needs to consider $w \gtrsim 0.65$ \cite{Haque:2022kez}, or $k=\frac{2+2w}{1-w} > 9$. This lower bound comes from the fact that, for higher $k$, the inflaton energy density redshifts faster (see Eq.~(\ref{rhophia})) so the transition to radiation domination is achieved sooner.  The requirement for large $w$ can be relaxed if one considers non-minimal couplings of the Higgs to the gravitational sector of the type \cite{CMOSV},
\beq
{\cal L}_\xi = -\frac{\xi}{2}|H|^2{\cal R},
\label{Eq:couplingricci}
\eeq
where ${\cal R}$ is the Ricci scalar. This
generates effective couplings between the inflaton and the Higgs boson
\beq
{\cal L}_\xi^{\phi H}=\frac{\xi}{M_P^2}
\left[2 V(\phi)-\frac{1}{2}g^{\mu \nu}\partial_\mu\phi \partial_\nu \phi\right] |H|^2.
\eeq
We can then write the Boltzmann equation for the radiation
\beq
\frac{\partial \rho_R}{\partial t} + 4 H \rho_R = R_k,
\label{Eq:friedmannrhor}
\eeq
where $R_k$ the amount of energy transferred per unit time and per unit volume. To compute $R_k$, one needs to add the standard
gravitational contribution corresponding to the exchange of graviton \cite{CMOV} and the non-minimal contribution from the coupling
(\ref{Eq:couplingricci}).

The rate can then be written $R_k = R_k^0 + R^\xi_k$ with
\begin{align}
R_k^0 & = N_s \frac{\rho_\phi^2}{16 \pi M_P^4}
\sum_{n=1}^\infty 2 n \omega|{\cal P}_{2n}^k|^2 \, ,
\\
R^\xi_k & =N_s \frac{2 \times \xi^2}{8 \pi M_P^4}
\sum_{n=1}^{\infty}2n\omega
\left|2{\cal P}_{2n}^k \rho_\phi
+\frac{(n \omega)^2}{2}\phi_0^2 |{\cal P}_n|^2 
\right|^2 , \nonumber
\end{align}
where $N_s=4$ is the number of real scalars in the Standard Model 
and we neglect the Higgs mass.
We considered even values of $k$, each mode $n$ transferring 
an energy $2n\omega$ per scattering to the bath.
For the case $k=2$, only the mode $n=1$ contributes, and 
we have (${\cal P}_1=\frac{1}{2}$, ${\cal P}_2^2=\frac{1}{4}$) which gives for $N_s=4$
\beq
R_2 = R_2^0 + R^\xi_2=\frac{m_\phi \rho_\phi^2}{32 \pi M_P^4}
(1+36 \xi^2),
\eeq
in agreement with \cite{CMOV} and \cite{CMOSV}. 
Taking into account the $\phi_0$ dependence in $\omega$ in Eq.~(\ref{eq:angfrequency}),
we can define
\beq
R_k \equiv\alpha_kM_P^5\left(\frac{\rho_\phi}{M_P^4}\right)^{\frac{5k-2}{2k}} .
 \eeq
The values of $\alpha_k$ are given in Table.~\ref{tab:alpha_k} for $k \le 20$.

\begin{table}[ht!]
\begin{tabular}{|l|l|}
\hline
$\alpha_{6\phantom{0}} = 0.000193 + 0.00766 \,\xi ^2$ 
    & $\alpha_{8\phantom{0}} = 0.000529 + 0.0205 \,\xi ^2$ \\ \hline
$\alpha_{10} = 0.000966 + 0.0367 \,\xi ^2$ 
    & $\alpha_{12} = 0.00144 + 0.0537 \,\xi ^2$ \\ \hline
$\alpha_{14} = 0.00192 + 0.0702 \,\xi ^2$ 
    & $\alpha_{16} = 0.00238 + 0.0855 \,\xi ^2$ \\ \hline
$\alpha_{18} = 0.00281 + 0.0993 \,\xi ^2$ 
    & $\alpha_{20} = 0.00319 + 0.112 \,\xi ^2$ \\ \hline
\end{tabular}
\caption{\small Coefficients $\alpha_k$ relevant for the rate of gravitational reheating.}
\label{tab:alpha_k}
\end{table}

Finally, we can solve Eq.~(\ref{Eq:friedmannrhor}) to obtain 
\begin{align}
\rho_R(a) & \simeq \alpha_k\frac{k+2}{8k-14}\sqrt{3}M_P^4
\left( \frac{\rhoe}{M_P^4}\right)^{\frac{2k-1}{k}}
\left(\frac{\ae}{a}\right)^4 , \\
\rhorh & = M_P^4\left(\frac{\rhoe}{M_P^4}\right)^{\frac{4k-7}{k-4}}
\left(\frac{\alpha_k \sqrt{3}(k+2)}{8k-14}\right)^{\frac{3k}{k-4}} \nonumber \\
& = \frac{g_*(\trh) \pi^2}{30} \trh^4 ,
\end{align}
which predicts the reheating temperature shown by the black dotted curves in Fig.~\ref{Fig:baryon_asymmetry_k} for different values of $\xi$, where the triangles along the curves correspond to even values of $k$. As can be seen, the black and colored curves are nearly parallel, and this correlation gives the upper x-axis of Fig.~\ref{Fig:baryon_asymmetry_mN} based on the value of $m_N$. The gray regions in Figs.~\ref{Fig:baryon_asymmetry_k} and \ref{Fig:baryon_asymmetry_mN} are inconsistent because even minimal gravitational interactions ($\xi = 0$) cannot achieve such low $\trh$.
That is, gravitational interactions alone provide a lower limit to the reheating temperature which is $k$-dependent. Furthermore, in much of the parameter space, gravitational interactions can provide sufficient reheating without the need of additional inflaton couplings.

{\bf \textit{Summary}}.---%
In this \emph{Letter}, we have demonstrated that purely gravitational interactions of the inflaton $\phi$ can produce a sufficient abundance of right-handed neutrinos, which later decay and generate the observed baryon asymmetry of the Universe via leptogenesis. This mechanism, which we call inflationary gravitational leptogenesis, can explain the baryon asymmetry for a wide range of right-handed neutrino masses $m_N$ and reheating temperatures $\trh$. This mechanism works for a class of inflationary models as long as the equation of state of $\phi$ is $w \ge 0.5$ during reheating, which is the case when the potential takes the form $\phi^k$ with $k \ge 6$ near the origin. Inflationary reheating can also successfully complete through gravitational interactions with the Higgs boson. This paradigm requires only gravitational interactions, so the results shown in Figs.~\ref{Fig:baryon_asymmetry_k} and \ref{Fig:baryon_asymmetry_mN} are robust and general. Indeed, these results provide a lower limit to the reheating temperature as gravitational interactions are necessarily present.  This paves the way for new opportunities in inflationary model building and baryogenesis.

{\bf \textit{Acknowledgements}}.---%
We would like to thank E.~Dudas, S.~Clery, and S.~Verner for useful discussions. This project has received support from the European Union's Horizon 2020 research and innovation programme under the Marie Sk$\lslash$odowska-Curie Grant Agreement No 860881-HIDDeN and the IN2P3 Master Projet UCMN. The work of R.C. and K.A.O.~was supported in part by DOE grant DE-SC0011842 at the University of Minnesota.

\end{document}